\documentclass[10pt,twocolumn,letterpaper]{article}

\usepackage{cvpr}              

\usepackage{graphicx}
\usepackage{amsmath}
\usepackage{amssymb}
\usepackage{booktabs}

\usepackage{array}
\usepackage{siunitx}
\usepackage[accsupp]{axessibility}  

\usepackage[pagebackref,breaklinks,colorlinks]{hyperref}

\usepackage{color}
\usepackage[capitalize]{cleveref}
\crefname{section}{Sec.}{Secs.}
\Crefname{section}{Section}{Sections}
\Crefname{table}{Table}{Tables}
\crefname{table}{Tab.}{Tabs.}

\newcommand{\added}[1]{{#1}}

\def\cvprPaperID{32} 
\def\confName{NTIRE}
\def\confYear{2022}

\definecolor{somegray}{rgb}{0.5, 0.5, 0.5}
\newcommand{\darkgrayed}[1]{\textcolor{somegray}{#1}}
\makeatletter
\newcommand*\titleheader[1]{\gdef\@titleheader{#1}}
\AtBeginDocument{%
  \let\st@red@title\@title
  \def\@title{%
    \vskip-3em
    \bgroup\normalfont\large\centering\@titleheader\par\egroup
    \vskip1.5em\st@red@title}
}

\makeatother

\titleheader{\darkgrayed{This paper has been accepted for publication at the\\
IEEE Conference on Computer Vision and Pattern Recognition Workshops (CVPRW), \\ 
New Orleans, 2022. \copyright IEEE}}

\input{floats/fig_eye_catcher}

\title{Multi-Bracket High Dynamic Range Imaging with Event Cameras}

\begin{document}

\author{Nico Messikommer\thanks{equal contribution}$^{\ ,1}\qquad $ 
Stamatios Georgoulis$^{*,2}\qquad$
Daniel Gehrig$^{1}\qquad$ 
Stepan Tulyakov$^{2}\qquad$ \\
Julius Erbach$^{2}\quad$ 
Alfredo Bochicchio$^{2}\quad$ 
Yuanyou Li$^{2}\quad$ 
Davide Scaramuzza$^{1}\quad$\\
\and
$^{1}$Dept. of Informatics, Univ. of Zurich and Dept. of Neuroinformatics, Univ. of Zurich and ETH Zurich \\
$^{2}$Huawei Technologies, Zurich Research Center
}
\newcommand{\NAME}{EHDR}
\maketitle

\begin{abstract}
\vspace{-4mm}
Modern high dynamic range (HDR) imaging pipelines align and fuse multiple low dynamic range (LDR) images captured at different exposure times. While these methods work well in static scenes, dynamic scenes remain a challenge since the LDR images still suffer from saturation and noise. In such scenarios, event cameras would be a valid complement, thanks to their higher temporal resolution and dynamic range. In this paper, we propose the first multi-bracket HDR pipeline combining a standard camera with an event camera. Our results show better overall robustness when using events, with improvements in PSNR by up to 5dB on synthetic data and up to 0.7dB on real-world data. We also introduce a new dataset containing bracketed LDR images with aligned events and HDR ground truth.

\end{abstract}

\vspace{-6mm}
\section*{Multimedia Material}
\vspace{-1mm}
Additional qualitative results can be viewed in this video: \url{https://youtu.be/fw9-gNg6cM8} 
\vspace{-2mm}
\section{Introduction}
\label{sec:introduction}

Natural scenes have considerable variations in their illumination. On a sunny day, the same scene may depict a bright sky or sun as well as deep shadows with a brightness ratio of 1:10,000. The human eye is accustomed to perceiving such a \textit{dynamic range} in natural scenes. Hence we expect the same from photos. However, conventional cameras have to set a global exposure time for the entire image and compress its full dynamic range into 10-14 bits. This is achieved through clipping, compression, and quantization of intensity values. As a result, captured images look less vivid and unimpressive. 
This problem is becoming more prominent as displays can natively support high dynamic range content.

\textit{Exposure bracketing}~\cite{Debevec97siggraph} is a popular method for acquiring \textit{high dynamic range~(HDR)} photos without special hardware. The method operates by capturing several \textit{low dynamic range~(LDR)} photos of the same scene under different exposures, aligning them, and fusing them together. This method provides great results when there is no camera or scene motion. Unfortunately, in the age of handheld smartphone photography, this solution has practical limitations. In the presence of scene or camera motion, this method must deal with LDR image misalignments and degradations. Several works~\cite{kalantari-et-al-2017,wu-et-al-2018,yan-et-al-2019_cvpr,yan-et-al-2020,lee-et-al-2020_ieee,liu-et-al-2021} tried to tackle these problems, but the latter cannot be robustly solved using standard techniques, \textit{e.g.} image-based alignment, because the bracketed LDR images violate the \textit{brightness constancy assumption}~\cite{tursun-et-al-2015,kalantari-et-al-2017}. \textit{Exposure compensation} on the bracketed LDR images is often used as a countermeasure, yet image saturation, noise, and motion blur still pose real challenges for these image-based HDR works.

By contrast, the human eye can reliably perceive a scene in a high dynamic range. 
Event cameras~\cite{brandli-et-al-2014,son-et-al-2017} are novel \emph{neuromorphic} vision sensors that attempt to mimic the high dynamic range and the high speed response of biological vision systems.
Instead of measuring synchronously absolute intensity frames at fixed time intervals, event cameras only measure the \emph{changes} in logarithmic intensity and do this independently for each pixel, resulting in an asynchronous stream of \emph{events}. 
The resulting data have high temporal resolution, HDR and do not suffer from motion blur. 
Recent works have leveraged the outstanding properties of event cameras to generate high-speed video reconstructions with HDR properties from events~\cite{wang-et-al-2019,rebecq-et-al-2019,zhang-et-al-2020,zou2021learning}.
Nonetheless, event cameras only measure changes in brightness, and thus global image reconstruction from events is ill-posed.
This places fundamental limits on these event-based HDR methods, which are further aggravated by persisting technical limitations of the current event sensor technology, \textit{i.e.}, low spatial resolution, and lack of events in low contrast regions.

To bypass these limitations, hybrid HDR works~\cite{han-et-al-2020,wang-et-al-2020} proposed to combine a single LDR image captured by a frame camera with events from an auxiliary event camera, thus leveraging the advantages of both.
The authors proposed different ways to enhance the luminance of the LDR image using the added information from events, but they still rely on the chrominance of the LDR image for color. 
Therefore, these methods have to hallucinate color in the areas where the LDR image is saturated. 
Moreover, since low contrast parts of the scene do not trigger events, these methods suffer from poor details and are highly dependent on motion since static scenes do not generate events.

In this paper, we propose to marry image-based exposure bracketing with event-based vision to get the best of both worlds. By doing so, we can enhance key parts of the HDR imaging pipeline, like the alignment of bracketed LDR images and their fusion to HDR at the feature level. Our main contributions can be summarized as:

\begin{enumerate}
    \item We introduce \NAME, the first method that combines bracketed LDR images and synchronized events for HDR imaging. In doing so, \NAME~is more robust than image-based HDR works to LDR image saturation, noise and motion blur, and, unlike event-based and hybrid HDR works, can faithfully reproduce color and low contrast details regardless of scene motion.
    \item We propose a deformable feature alignment module that leverages motion information from both images and events to guide the learning of kernel offsets and modulation masks, which, in turn, leads to significant performance improvements in PSNR and SSIM on both synthetic and real data. 
    \item To evaluate the proposed method and facilitate future research on the topic, we collect the first HDR dataset consisting of sequences with ground truth HDR, bracketed LDR images, and synchronized and aligned event data, called HDR-ERGB. 
\end{enumerate}

\section{Related Work}
\label{sec:related_work} 

Current HDR solutions can be classified into: single-exposure and multi-exposure methods that only rely on standard image sensors; event-based methods that only use neuromorphic sensors; and hybrid methods that utilize both.
For a more detailed overview, we refer to~\cite{wang21pami}

\textbf{Single-exposure} methods infer an HDR image~\cite{rempel-et-al-2007,eilertsen-et-al-2017,marnerides-et-al-2018,liu-et-al-2020,santos-et-al-2020} or a stack of differently exposed LDR images~\cite{endo-et-al-2017,lee-et-al-2018} from a single LDR image, and can therefore be applied to legacy LDR content. They operate by inverting \textit{non-linear mapping}, \textit{quantization} and \textit{saturation clipping} applied during image acquisition; thus, they are also called \textit{inverse tone mapping~(iTM)} methods. Deep learning iTM approaches~\cite{endo-et-al-2017,eilertsen-et-al-2017,marnerides-et-al-2018,liu-et-al-2020,santos-et-al-2020} recently achieved good results in recovering saturated details by utilizing large image context; however, they still essentially hallucinate details in saturated regions from the surrounding non-saturated context, and, thus, are not suitable for commercial applications. 

Other works~\cite{cai-et-al-2018,jiang-et-al-2019,guo-2020-et-al,yang-et-al-2020} expose hidden details in dark areas by tone mapping; however, they are also not able to recover details that are not present in the original image.  

\textbf{Multi-exposure} HDR methods acquire multiple LDR images under different exposures and fuse them into an HDR image. The LDR images with different exposures can be acquired at once using a \textit{special camera system}, \textit{e.g.} system with two image sensors and common lens with beam splitter~\cite{tocci-et-al-2011} or two separate cameras~\cite{prabhakar-et-al-2017,trinidad-et-al-2019}. However, such systems have higher power consumption and memory requirements. There exist more exotic systems too, such as systems with pixel-wise exposure control~\cite{serrano-2016-et-al}, out-of-range intensity warping~\cite{zhao-et-al-2015}, or gradient encoding~\cite{tumblin-et-al-2005}, but the latter are expensive and inaccessible to regular users.      

Alternatively, LDR images can be acquired by standard camera hardware using \textit{exposure bracketing} as in~\cite{sen-et-al-2012,kalantari-et-al-2017,wu-et-al-2018,yan-et-al-2019_cvpr,yan-et-al-2019_springer,ma-et-al-2019,ou-et-al-2020,liu-et-al-2021,niu2021hdr-gan,Yan2021accurateHDR,Pu2020robustHDR,Choi2020pyramid_hdr} by taking several photos of the same scene under different exposures. However, this technique has a longer capturing time, which, in turn, can result in misaligned and blurry LDR images for longer exposures due to camera motion or non-static scenes. The latter can be avoided by synthetic exposure, \textit{i.e.} capturing multiple underexposed images with a high gain and synthesizing long exposed images~\cite{hasinoff-et-al-2016,liba-et-al-2019}. Still, synthetic exposure provides limited dynamic range improvement and still suffers from the images' misalignment problem. Note that, a naive fusion of the misaligned LDR images is prone to ghosting artifacts in the HDR image. 

\textbf{Multi-exposure alignment}~(see survey~\cite{tursun-et-al-2015}). A simple solution to the alignment problem is to \textit{reject moving object pixels}~\cite{oh-et-al-2014,yan-et-al-2019_cvpr} or use robust exposure fusion~\cite{hasinoff-et-al-2016,liba-et-al-2019}. However, these methods often fail to identify moving objects and are unable to reconstruct them in HDR. Another solution is to explicitly \textit{estimate and compensate motion} between LDR images~\cite{hacohen-et-al-2011,gallo-et-al-2015}. These methods register all exposures to a reference LDR image by estimating global transformations~(e.g., homography) or optical flow. Since different exposures have intensity differences, motion estimation is performed using robust similarity measures~\cite{hasinoff-et-al-2016}, or features~\cite{gallo-et-al-2015}, or is performed after exposure compensation~\cite{kalantari-et-al-2017}. Still, it is hard to directly estimate motion from bracketed LDR images because long exposures suffer from saturation and motion blur, while short exposures from image noise. Also, image-based motion estimation methods suffer from ambiguity when the motion is large. 

\textbf{Neuromorphic sensor}. All HDR approaches described so far rely on frame-based cameras that capture the entire image at once. However, the dynamic range of frame-based cameras is limited as they need to encode all scene intensities into a fixed number of bits (e.g., 8 bits). While frame-based cameras are the sensor of choice for computer vision applications, new biologically-inspired \textit{neuromorphic sensors}~\cite{Gallego20pami} are gaining popularity. Instead of measuring the intensity of every pixel, these sensors report asynchronous events whenever the log-intensity change at an individual pixel reaches a certain threshold, called \textit{contrast threshold}. Since neuromorphic sensors do not encode absolute intensity levels, they do not saturate in extreme lighting conditions, such as bright daylight and night, and they currently have $>120$ dB dynamic range versus $50$ dB - $80$ dB for a frame camera. Because of these properties, event cameras have recently become popular for HDR reconstruction. 
 
\textbf{Event-based} methods~\cite{rebecq-et-al-2019,wang-et-al-2019,zhang-et-al-2020,zou2021learning} reconstruct intensity frames from events using deep learning. These methods do not explicitly target HDR imaging, but they rather reconstruct HDR-like images as a consequence of using events. Their reconstruction quality is typically low in part because current event cameras have low spatial resolution~($\leq 1$ MP) and do not provide events in low contrast regions. These problems may be alleviated in the future through the advancement of the current sensor technology. Yet, the resolution of event cameras will always lag behind frame-based cameras due to the higher complexity of their pixel circuitry. Still, there are more fundamental limitations. First, the reconstruction of an image from events is an ill-posed problem due to the lack of absolute intensity information and varying contrast thresholds. Therefore, event-based methods often produce images with incorrect global contrast. Second, event cameras are unable to capture details below the contrast threshold, which becomes evident in low light as noted in~\cite{zhang-et-al-2020}. Therefore, the results of purely event-based methods typically lack details. Finally, the quality of event-based image reconstruction is motion-dependent, usually performing poorly on scenes with little motion.

\textbf{Hybrid} single-exposure HDR methods~\cite{han-et-al-2020,wang-et-al-2020} combine an LDR image captured by a high-resolution frame camera with events acquired by an auxiliary event camera. These methods propose different ways to enhance the luminance of the LDR image using the added information from events. Yet, they still rely on the chrominance of the LDR image when it comes to color, as most event cameras currently do not provide color information. Inevitably, in saturated or underexposed regions, these approaches ought to hallucinate color, and sometimes even structure, from the non-saturated nearby context, which is unreliable when the saturated regions are large. Essentially, current single-exposure hybrid works suffer the same drawbacks as iTM methods, that is, they hallucinate results, yet they produce more educated guesses guided by the added event information. In contrast, we propose the \textit{first multi-exposure hybrid method} that eliminates any hallucination component and instead relies on actual measurements generated by combining exposure bracketing with event-based vision.

\section{Method}
\label{sec:method}

\newcommand{\Interval}[3]{$#1_{\scriptscriptstyle [#2, #3)}$}
\newcommand{\ArrowChunks}[1]{(..., $#1_{0 \rightarrow -\tau}$, $#1_{0\rightarrow \tau}$, $#1_{\tau \rightarrow 2\tau}$, ...)}
\newcommand{\IntervalChunks}[1]{($#1_{ \scriptscriptstyle [0, \tau]}$, $#1_{\scriptscriptstyle [\tau, 2\tau]}$, ...)}

\subsection{Method Overview}
\label{sec:method_overview}

\begin{figure*}
\centering
\includegraphics[width=0.8\textwidth]{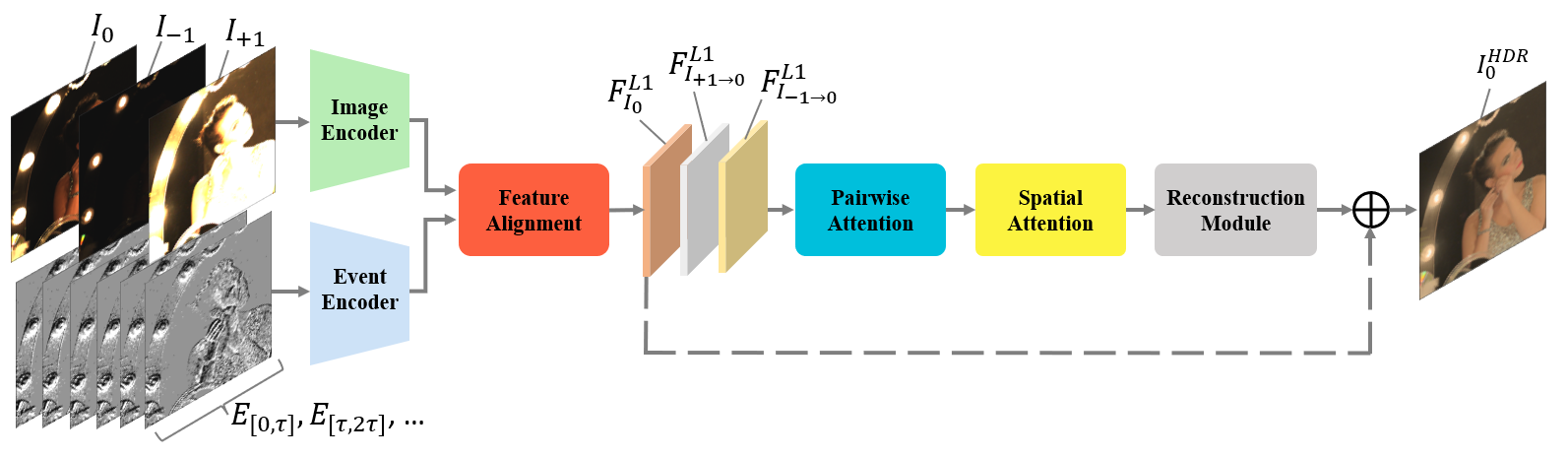}
\vspace{-2mm}
\caption{Overview of our proposed multi-bracket HDR pipeline. The events and frames are used to generate aligned features, which are processed by a pairwise and a spatial attention module. Finally, the reconstruction module outputs the HDR prediction.}\label{fig:meth_overview}
\vspace{-5mm}
\end{figure*}

Let us assume the following setup. An image sensor captures a burst of bracketed LDR images ($I_{0}$, $I_{-1}$, $I_{+1}$) corresponding to mid, short and long exposure times, respectively. The images are captured by a handheld camera and have significant camera and scene motion between them. In parallel, an event sensor records a stream of asynchronous events $E=((t_0,x_0,y_0,p_0),(t_1,x_1,y_1,p_1),\dots) $, with $t$ denoting the time that an event was triggered, $(x,y)$ the spatial position of the event, and $p$ its polarity (positive or negative). We assume that the event data and the images are temporally synchronized and spatially aligned and have the same resolution, as though as they come from one sensor (\textit{e.g.} a hybrid sensor). Our goal is to reconstruct the HDR image at mid exposure time $I_{0}^{HDR}$.

Fig.~\ref{fig:meth_overview} gives an overview of our method, called \NAME. 
As a pre-processing step, the stream of raw events $E$ is split into fixed-duration chunks \IntervalChunks{E}, with each chunk containing all events within a time window $\tau$.
\added{The events within each chunk are represented as a \textit{voxel grid}~\cite{zihao-et-al-2018,Gehrig19iccv} with 5 equally-sized temporal bins.}
The bracketed LDR images and chunked events are passed to an \textit{Image Encoder} and \textit{Event Encoder}, respectively.
The resulting multi-scale feature representations of each non-reference image are aligned to the reference image $I_{0}$ by a \textit{Feature Alignment} module using the chunked event features. 
In a next step, the aligned feature representations ($F_{I_{-1 \rightarrow 0}}^{L1}$, $F_{I_{+1 \rightarrow 0}}^{L1}$) and the reference features ($F_{I_{0}}^{L1}$) are concatenated and inputted to a \textit{Pairwise Attention} module that fuses them into a single feature representation that is consequently passed to a \textit{Spatial Attention} module that aims to recover fine details by deep spatial feature transform.
Finally, the attended feature representation is decoded by a \textit{Reconstruction} module that produces the HDR image prediction $\hat{I}_{0}^{HDR}$. 
Below, we explain the individual modules of \NAME~in more detail.

\subsection{Encoder \& Reconstruction Modules}
\label{sec:feature_encoder_decoder}

Our system consists of two encoders, \textit{i.e.}, \textit{Image Encoder} and \textit{Event Encoder}, that follow the same multi-scale architecture. 
In particular, each encoder uses 5 residual blocks~\cite{He16cvpr}, each having a (Conv2d, ReLU, Conv2d) design with a residual connection. 
Unless stated otherwise, all convolutions use 3x3 kernels with 64 channels, padding 1, stride 1, and no BatchNorm layers. 
The 5 residual blocks are followed by 2 down-sampling blocks, each with a (Conv2d, LeakyReLU, Conv2d, LeakyReLU) design with the first Conv2d having a stride of 2, essentially generating a pyramid of feature representations across 3 scales (L1: 1, L2: 1/2, L3: 1/4). 
The multi-scale architecture allows for coarse to fine feature alignment inside the \textit{Feature Alignment} module, such that larger motions can be compensated too. 
Note that, the input to the \textit{Event Encoder} module are chunked events in voxel grid format with 5 channels, while the \textit{Image Encoder} module expects images with 6 channels, as in prior HDR works~\cite{kalantari-et-al-2017,kalantari2019deep,yan-et-al-2019_cvpr}. 
In particular, the non-linear LDR image $I_{j}$ is concatenated with its exposure compensated linearized version $\tilde{I}_{j} = f^{-1}(I_{j})/t_{j}$, where $f$ is the camera response function, simplified to a gamma curve with $\gamma=2.2$ in our case\footnote{We assume that a proper image linearization has been performed in advance using the camera response function (CRF) computed from camera calibration techniques. Hence, gamma can replace CRF in this case.}, and $t_{j}$ is the exposure time of image $j$. 
This exposure compensation procedure approximates brightness constancy among the LDR images required for alignment purposes.

The \text{Reconstruction} module consists of 10 residual blocks that follow the exact same design as the encoder ones. Note that, a skip connection from the feature representation of the reference LDR image $F_{I_{0}}^{L1}$ is added to the output of this module before decoding it to the HDR image $\hat{I}_{0}^{HDR}$ using a Conv2d layer (64 to 3 channels).

\subsection{Feature Alignment Module}
\label{sec:feature_alignment}

Exposure bracketing in handheld photography of dynamic scenes may lead to image misalignments, which need to be resolved in order to avoid ghosting artifacts in the HDR image. 
Existing HDR works estimate some form of 'motion', be it global or local, between the LDR images after exposure compensation~\cite{kalantari-et-al-2017,wu-et-al-2018,yan-et-al-2019_cvpr,yan-et-al-2020,lee-et-al-2020_ieee,liu-et-al-2021}. 
The latter is implicitly required for motion estimation which is dependent on brightness constancy in the LDR images. 
However, even exposure compensation can not guarantee brightness constancy, as the LDR images also suffer from saturation, noise, or motion blur. 
This renders motion estimation from bracketed LDR images an ill-posed problem. Events do not suffer from saturation and motion blur, and carry fine-grained information about motion between bracketed LDR images due to their high temporal resolution. 
As a result, they are a natural fit for image alignment purposes in dynamic scenes, relaxing the strong assumptions of brightness constancy and motion linearity in current HDR works. 
However, events are sparse by nature and absent in low contrast regions due to the limited contrast sensitivity of current event cameras, rendering them incomplete for motion estimation in every image patch if used on their own. 
To leverage this complementarity between images and events and get the best of both worlds, in this paper, we propose to combine these two sources of motion information for LDR image alignment in HDR photography.

\begin{figure}
\centering
\includegraphics[width=0.48\textwidth]{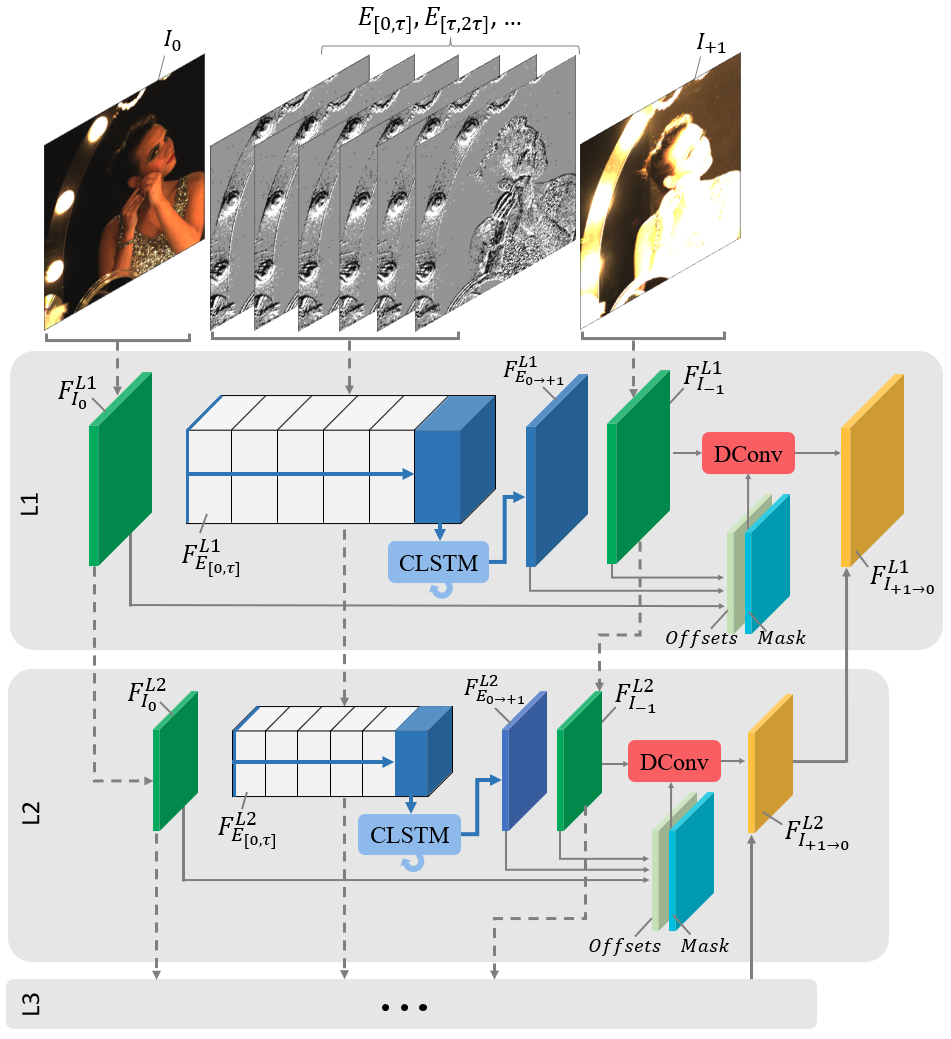}
\vspace{-7mm}
\caption{Overview of our alignment module. The event features (blue) are processed by a ConvLSTM (CSTM) and afterwards combined with image features (green) to compute the offsets and masks for the deformable convolution (DConv). The output of the deformable convolutions are the aligned image features (yellow).}\label{fig:align_module}
\vspace{-5mm}
\end{figure} 

To this end, we design a \textit{Feature Alignment} module that leverages information from both images and events to align the non-reference images at the feature level. Our starting point is the Pyramid, Cascading, and Deformable (PCD) module introduced in EDVR~\cite{wang2019edvr}, used in HDR too~\cite{liu-et-al-2021}. The PCD module uses a pyramid of feature representations to estimate kernel offsets and modulation masks for modulated deformable convolutions~\cite{zhu2019deformable}
\footnote{The modulated deformable convolution is computed as $y(p) = \sum_{k=1}^{K} w_{k} \cdot x(p + p_{k} + \Delta p_{k}) \cdot \Delta m_{k}$, with $\Delta p_{k} \in \mathbb{R}$ and $\Delta m_{k} \in [0, 1]$ being the learnable kernel offsets and modulation masks respectively.} 
at each pyramidal level separately, which are then used to gradually 'align' the feature representations of non-reference images to the reference image. We build upon this core idea of deformable feature alignment, but introduce the following modifications: 

(1) The kernel offsets and modulation masks at level $L_{l}, l \in \{1,2,3\}$ are jointly computed from image ($F_{I_{0}}^{L_{l}}$, $F_{I_{+1}}^{L_{l}}$) and chunked event ($F_{E_{[0, \tau]}}^{L_{l}}$, $F_{E_{[\tau, 2\tau]}}^{L_{l}}$, ..., $F_{E_{[-\tau, +1]}}^{L_{l}}$) features. Fig.~\ref{fig:align_module} illustrates this procedure. Although image features can be directly used for the computation, chunked event features first need to be integrated across the entire time window ($\{0, \tau, ..., +1\}$). We achieve this by introducing ConvLSTM modules~\cite{xingjian2015convolutional}, one at each level $L_{l}$, that output the integrated event features $F_{E_{0 \rightarrow +1}}^{L_{l}}$. An added benefit of the ConvLSTM modules is that they can 'compensate' the camera motion in-between chunked events, which inevitably causes events to appear in different locations than the one they were originally triggered. The integrated event features $F_{E_{0 \rightarrow +1}}^{L_{l}}$ are concatenated with the image features ($F_{I_{0}}^{L_{l}}$, $F_{I_{+1}}^{L_{l}}$) and passed through a (3 x (Conv2d, LeakyReLU), Conv2d) block that returns the offsets and masks of each level $l$. Note that, we described the procedure for time window ($\{0, \tau, ..., +1\}$), but the exact same holds for time window ($\{0, \tau, ..., -1\}$)

(2) The kernel offsets for level $l$ are learned as a residual to the kernel offsets estimated at level $l-1$. By doing so, we encourage a coarse-to-fine learning of motion that takes into account estimates from the low-resolution feature representations with larger receptive field. Thus, allowing for compensation of larger motions. 

\subsection{Attention Modules}
\label{sec:feature_attention}

After the \textit{Feature Alignment} module, the resulting feature representations ($F_{I_{0}}^{L1}$, $F_{I_{-1 \rightarrow 0}}^{L1}$, $F_{I_{+1 \rightarrow 0}}^{L1}$) are passed through a pair of consecutive attention modules. 

First, \textit{Pairwise Attention} aims to fuse information from the different LDR images. To design it, we draw inspiration from the HDR generation procedure of ground truth data. In case of no camera or scene motion, an HDR image can be approximated via simple weighted averaging of the exposure compensated LDR images~\cite{kalantari-et-al-2017}
\footnote{For the case of 3 bracketed LDR images ($I_{j}$, $j \in [0, -1, +1]$), the HDR image can be computed as $I_{0}^{HDR}(p) = \frac{\sum_{j} \alpha_{j}(p) \tilde{I}_{j}(p)}{\sum_{j} \alpha_{j}(p)}$, where $\tilde{I}_{j} = I_{j}^{\gamma}/t_{j}$. Here, $\alpha_{j}(p)$ represent per-pixel blending weights.}. 
Assuming that motion has been compensated in the previous module, we extend this concept to the feature level. In particular, we use attention blocks with a (Conv2d, LeakyReLU, Conv2d, Sigmoid) design each, that are applied pairwise between ($F_{I_{0}}^{L1}$, $F_{I_{-1 \rightarrow 0}}^{L1}$, $F_{I_{+1 \rightarrow 0}}^{L1}$) and $F_{I_{0}}^{L1}$, and provide per-pixel and per-channel blending weights. The latter are used as guides for weighted averaging in the feature level, resulting in a single merged feature representation.

Next, \textit{Spatial Attention} aims to recover fine details from the merged feature representation. For this, we directly utilize the multi-scale spatial attention from the TSA module in EDVR~\cite{wang2019edvr}.

\subsection{HDR-ERGB Dataset}
\label{sec:hdr_dataset}

As there is no publicly available HDR dataset which features RGB images and synchronized event data, we captured a new dataset named HDR Events and RGB (HDR-ERGB) dataset. 
The hybrid imaging system used to capture our dataset combines a high-resolution RGB camera synchronized with a high-resolution event sensor. 
The RGB camera is a FLIR Blackfly S with a resolution of 4000$\times$3000 and global shutter. 
The event camera is a Prophesee Gen4 with a resolution of 1280$\times$720. 
The two sensors share a similar FOV and are mounted in a beam splitter setup, containing a mirror which splits the incoming light to the event and frame camera, and ensures alignment between events and high-resolution frames.

To record bracketed LDR images with events and HDR ground truth, we divide the recording procedure into two steps. 
In step 1, we acquire HDR ground truth by mounting the imaging system on a tripod and recording a set of 9 bracketed LDR images (0, $\pm$ 1, $\pm$ 2, $\pm$ 4 fstops) with no camera or scene motion. 
In step 2, we acquire the bracketed LDR images (0, $\pm$ 2 fstops) and synchronized events with camera and/or scene motion. 
To simulate dynamic scenes, we follow a procedure similar to~\cite{kalantari-et-al-2017}, and capture camera motion by simply moving the tripod and scene motion by asking people to move. 
That is, the persons stand still during HDR ground truth recording (step 1), and receive an audio signal to start moving after the reference bracketed LDR image (0 fstops) is taken (step 2). 
Note that, a stream of events from the event camera is synchronously recorded with the bracketed LDR images at the beginning of step 2. 
Following this procedure, the reference bracketed LDR image of the motion affected recording (step 2) is aligned with the ground truth HDR recording (step 1). 
Stereo alignment between events and bracketed LDR images is performed by camera calibration and rectification, which results in events and images with a resolution of 960$\times$688. In total, we have collected 53 dynamic scenes (scene motion, with or without camera motion) and 12 static scenes (only camera motion). 
For more details, visit the supplementary materials.

\newcolumntype{C}[1]{>{\centering}m{#1}}

\section{Results}

\subsection{Experimental Settings}
\label{sec:settings}

\textbf{HDR dataset with synthetic events.} We use the HDM-HDR-2014 dataset~\cite{froehlich-2014-spie} that contains real-world HDR video sequences, which, in turn, can be used to synthesize events. To generate bracketed LDR images ($0\pm$3 f-stops) with realistic noise characteristics from the HDR video sequences, we follow the exact same procedure as in~\cite{kalantari2019deep}, with the exception of not applying tone perpetuation. To simulate the high-speed nature of event cameras, we use a frame skip of 2 between bracketed LDR images. Synchronized events are generated using the VID2E simulator~\cite{Gehrig20cvpr}. Following~\cite{stoffregen-2020-eccv}, we set the contrast threshold in VID2E to match the event rate per frame of HDR-ERGB, in order to ensure realistic event data rates. 
The HDM-HDR-2014 dataset was also used in the NTIRE 2021 Multi-Frame HDR Challenge~\cite{perez-pellitero-2021-cvpr}. However, we can not use the challenge dataset, as we lack access to the test set images required to synthesize events.

\newcommand\cellwidthhdm{1.35cm}

\begin{table}
\caption{Quantitative comparison with state-of-the-art multi-bracket HDR approaches on the synthetic HDM-HDR-2014.}
\vspace{-2mm}
\centering
\scalebox{0.9}{
\begin{tabular}{m{2.25cm}C{1.4cm}C{\cellwidthhdm}C{1.1cm}>{\centering\arraybackslash}m{1.2cm}}
  & \multicolumn{4}{c}{$\pm$3 f-stops} \\
 \cmidrule(lr){2-5}
Method  & PSNR-$\mu$$\uparrow$ & SSIM-$\mu$$\uparrow$ & LPIPS$\downarrow$ & HDR-VDP2$\uparrow$  \\
 \hline
Kalantari~\cite{kalantari-et-al-2017} & 39.53 & 98.21 & 0.0310 & 45.33  \\
AHDR~\cite{yan-et-al-2019_cvpr}       & 39.70 & 98.49 & 0.0230 & 47.52  \\
ADNet~\cite{liu-et-al-2021}           & 40.14 & 98.79 & 0.0222 & 47.37  \\
Ours w/o events                       & 40.42 & 98.67 & 0.0211 & 47.38  \\
 \hline
Ours                                  & \textbf{45.86} & \textbf{98.88} & \textbf{0.0161} & \textbf{53.21} \\
\end{tabular}
}
\label{tab:exp_hdm-hdr}
\end{table}

\begin{figure}[h]
\centering
\includegraphics[width=0.38\textwidth]{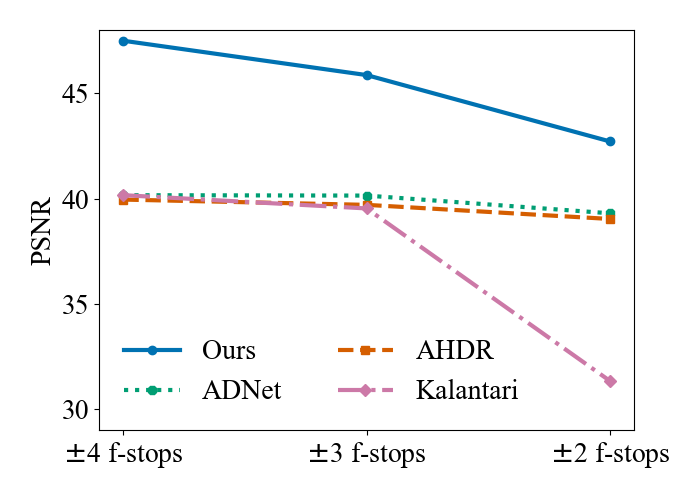}
\vspace{-4mm}
\caption{Qualitative experiments showing the effect of changing the f-stop values.}\label{fig:fstop}
\vspace{-4mm}
\end{figure} 
\begin{figure}[h]
\centering
\includegraphics[width=0.47\textwidth]{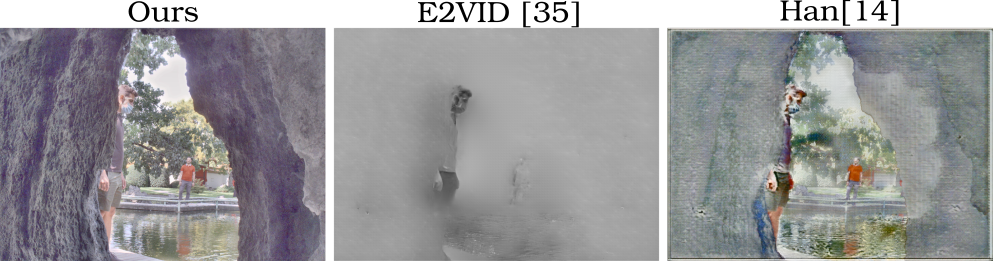}
\vspace{-2.5mm}
\caption{The comparison with existing event-based HDR method.}

\label{fig:event_comparison}
\end{figure} 
\newcommand\cellwidthreal{1.5cm}

\begin{table*}
\caption{Quantitative comparison with state-of-the-art multi-bracket HDR approaches on our recorded HDR-ERGB.}
\vspace{-3mm}
\centering
\scalebox{0.9}{
\begin{tabular}{m{2.4cm}C{\cellwidthreal}C{\cellwidthreal}C{\cellwidthreal}C{\cellwidthreal}C{\cellwidthreal}C{\cellwidthreal}C{\cellwidthreal}>{\centering\arraybackslash}m{\cellwidthreal}}
  & \multicolumn{4}{c}{$\pm$4 f-stops GT} & \multicolumn{4}{c}{$\pm$2 f-stops GT} \\
 \cmidrule(lr){2-5} \cmidrule(lr){6-9}
Method  & PSNR-$\mu$$\uparrow$ & SSIM-$\mu$$\uparrow$ & LPIPS$\downarrow$ & HDR-VDP2$\uparrow$ & PSNR-$\mu$$\uparrow$ & SSIM-$\mu$$\uparrow$ & LPIPS$\downarrow$ & HDR-VDP2$\uparrow$  \\
 \hline
Kalantari~\cite{kalantari-et-al-2017} & 31.15 & 84.75 & 0.130 & 48.14 & 31.94 & 85.01 & 0.124 & 49.39  \\
AHDR~\cite{yan-et-al-2019_cvpr}       & 31.98 & 86.79 & 0.132 & 50.97 & 32.53 & 87.20 & 0.127 & 53.25  \\
ADNet~\cite{liu-et-al-2021}           & 31.91 & 86.87 & 0.131 & 49.90 & 32.59 & 87.21 & 0.127 & \textbf{53.54}  \\
Wu ~\cite{wu-et-al-2018}              & 32.16 & 86.47 & 0.132 & 50.18 & 32.73 & 86.85 & 0.127 & 51.64  \\
 \hline
Ours                                  & \textbf{32.84} & \textbf{87.84} & \textbf{0.112} & \textbf{51.86}  & \textbf{33.32} & \textbf{88.22} & \textbf{0.108} & 53.32  \\
\end{tabular}
}
\label{tab:real}
\vspace{-3mm}
\end{table*}

\begin{figure*}[h]
\centering
\includegraphics[width=0.97\textwidth]{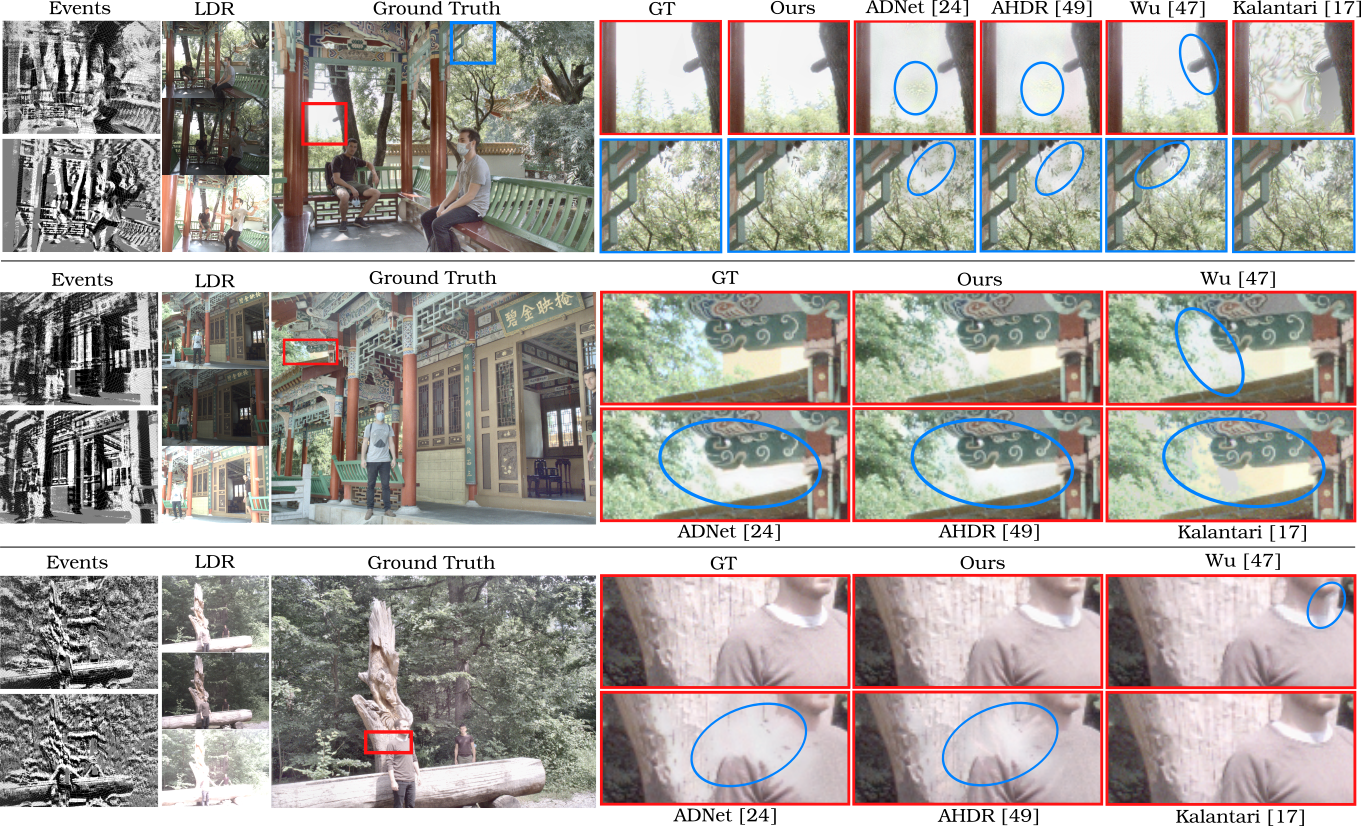}
\vspace{-3mm}
\caption{Our method can reliably align the LDR images without generating artifacts and can reconstruct thin structures e.g. leaves on a tree.
In comparison, the baselines suffer from misalignment artifacts and have difficulties reconstructing thin details. Failure cases are highlighted with a blue circle.}
\label{fig:real_overall}
\vspace{-3.5mm}
\end{figure*} 
\begin{figure*}[h]
\centering
\includegraphics[width=0.97\textwidth]{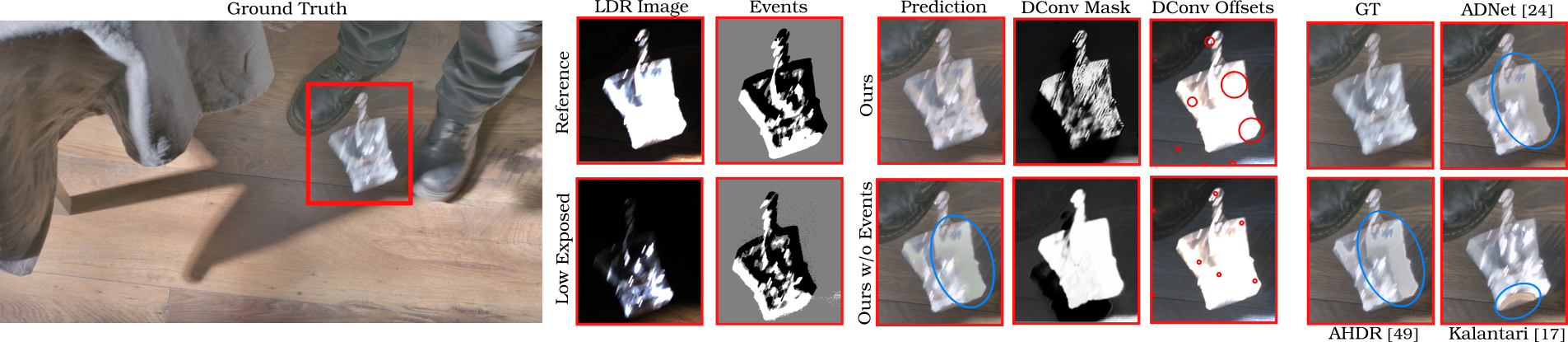}
\vspace{-3mm}
\caption{Qualitative example of the HDM-HDR 2014 dataset showing a falling glass. It can be observed that events provide more reliable motion information under fast motion than the different exposed LDR images. The comparison to the image-based version of our method shows a better HDR reconstruction due to a more detaild deformable modulation mask (DConv Mask) and larger deformable offsets (DConv Offsets). Finally, our method constructs a more accurate HDR image than the evaluated state-of-the-art baselines.}\label{fig:glass}
\vspace{-5mm}
\end{figure*} 
\begin{figure}[h]
\centering
\includegraphics[width=0.47\textwidth]{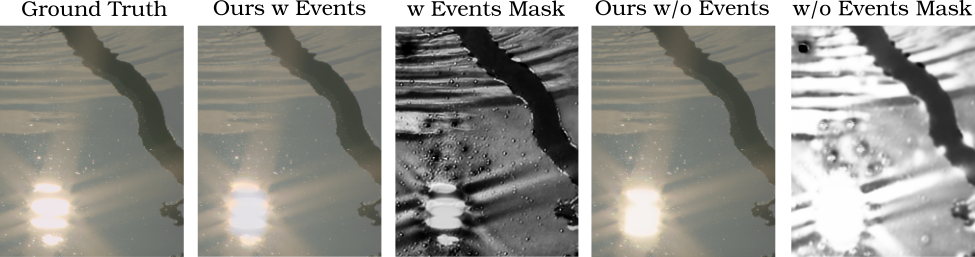}
\vspace{-2.5mm}
\caption{
The impact of the high-dynamic-range in events.
}
\vspace{-5.0mm}
\label{fig:wo_events}
\end{figure} 
\begin{table}[bp]
\centering
\caption{
Network ablation study on HDM-HDR-2014.
}
\vspace{-3mm}
\scalebox{0.9}{
\begin{tabular}{m{1.4cm}C{0.9cm}C{1.55cm}C{1.4cm}>{\centering\arraybackslash}m{1.5cm}}
Method  & Ours & w/o \newline feat. align. & w/o temp. att. & w/o \newline spat. att.\\

\cmidrule(lr){1-5}
PSNR-$\mu$$\uparrow$ & \textbf{45.86} & 40.37 & 45.50 & 45.32
\end{tabular}
}
\label{tab:hdm-hdr2014_ablations_short}
\end{table}

\textbf{Baselines.} 
As we propose the first multi-bracket HDR method with events, we are restricted to comparing against image-based multi-bracket HDR works. 
From the latter, we select the winner of the NTIRE 2021 Multi-Frame HDR Challenge~\cite{perez-pellitero-2021-cvpr} and the current state-of-the art method called ADNet~\cite{liu-et-al-2021}, the HDR method of Kalantari et el.~\cite{kalantari-et-al-2017}, and AHDR~\cite{yan-et-al-2019_cvpr}. 
Additionally, we include the method of Wu et al.~\cite{wu-et-al-2018} for comparison on HDR-ERGB only, since we could not train it with reasonable performance on the synthetic dataset.
Existing event-based methods~\cite{Rebecq19pami, han-et-al-2020} naturally perform worse, see Fig.~\ref{fig:event_comparison}, since they tackle a harder task, where only events and a single LDR image are given as input.
Thus, we did not include them in our comparisons

\textbf{Training details.} For supervision, we use a combination of L1 and LPIPS~\cite{Zhang18cvprLPIPS} losses, with weights $1$, on the $\mu$-law HDR images. $\mu$-law is a compression introduced in~\cite{kalantari-et-al-2017}, defined as $T = \text{log}(1 + \mu H) / \text{log}(1 + \mu)$ with $\mu = 5000$, that simulates a differentiable tonemapping operation. The following random augmentations are applied: scale, crop (256x256 patches), rotate (90 degrees), flip (horizontal and vertical), and color channel swap. We a use batch size of 4, Adam optimizer with an initial learning rate of $1 \times 10^{-4}$ that decreases by a factor of 2 every 15 epochs (300 epochs on HDR-ERGB), and train for 60 epochs (1500 epochs on HDR-ERGB). To guarantee a fair comparison, we trained all baselines with the same settings from scratch on both HDM-HDR-2014 and HDR-ERGB. 

\textbf{Testing details.} In HDM-HDR-2014, we test on 5 sequences (bistro\_03, carousel\_fireworks\_09, fireplace\_02, fishing\_closeshot, poker\_fullshot). We evaluate at the original image resolution with a cropped border of 10 pixels, \textit{i.e.} 1900$\times$1060, which is applied to remove the black pixels at the border. In HDR-ERGB, we test on 9 challenging sequences containing real noise for both images and events. Since we recorded ground truth HDR with up to $\pm$4 f-stops but bracketed LDR with $0\pm$2 f-stops, we can evaluate all methods on $\pm$2 f-stop or $\pm$4 f-stop range. The latter can test the hallucination capabilities of all methods outside the recorded dynamic range.

\textbf{Metrics.} We report results for the LPIPS, PSNR, SSIM and HDR-VDP-2~\cite{mantiuk-et-al-2011-tog} metrics.
As commonly done, we compute all metrics on the tonemapped images using $\mu$-law (-$\mu$) except HDR-VDP-2, \added{which is calculated using linear HDR images and default parameters (PPD: 52.72).}

\subsection{Comparison on HDM-HDR-2014}
\label{sec:comparison_synth}
We first evaluate all methods on the HDM-HDR-2014 dataset containing synthesized events.
The results in Table~\ref{tab:exp_hdm-hdr} verify the substantial benefit of including events in a multi-bracket HDR approach.
Our method significantly outperforms the state-of-the-art baselines on all the evaluated metrics.
To showcase the advantages of our method, an example containing a fast-moving object overexposed in the reference frame is shown in Fig.~\ref{fig:glass}.
Using the high-speed and high dynamic range events, our method can take the necessary information from the short exposure and reconstruct the glass without major artifacts.
The image-based baselines fail at properly aligning the moving object to its saturated counterpart in the reference frame.

Synthetic evaluation enables us to test all methods under different f-stops for the LDR brackets, which effectively increases or decreases the dynamic range recorded in the input LDRs.
The plot in Fig.~\ref{fig:fstop} illustrates the achieved PSNR score of the methods for the different f-stops.
It can be observed that our method shows higher performance with only $\pm$ 2 f-stops compared to all baselines at the larger range of $\pm$ 4 f-stops.
This shows the potential of our method to reduce the acquisition time for multi-bracket HDR, minimizing the risk for LDR image misalignments. We refer to Table 1 in the supplement for the numerical results of this comparison.

\subsection{Comparison on HDR-ERGB}
\label{sec:comparison_real}
To test under more realistic conditions, we evaluate all methods on our challenging HDR-ERGB Dataset.
We report the results for ground truth construction with different dynamic ranges in Table~\ref{tab:real}.
Our method outperforms the state-of-the-art baselines in all metrics, except for HDR-VDP2 on the $\pm2$ f-stop ground truth.
In general, the comparison on the $\pm2$ f-stop ground truth is more challenging for our method since the network needs to decide which events should be discarded as events contain a higher dynamic range than the ground truth image.
\added{
Compared to the synthetic HDM-HDR-2014 dataset, the events in our HDR-ERGB dataset contain real sensor noise, which can explain the lower performance improvement of \NAME~on real data.
}

The qualitative results on HDR-ERGB validate the advantages of \NAME. 
Since we use events and images, our method achieves a better LDR alignment compared to pure image-based alignment methods~\cite{kalantari-et-al-2017, wu-et-al-2018}, which exhibit ghosting artifacts due to misalignment, see Fig.~\ref{fig:real_overall} top.
Moreover, the image-based flow can fail in textureless regions, which leads to severe artifacts in the HDR prediction for~\cite{kalantari-et-al-2017}.
HDR methods relying on deep alignment~\cite{yan-et-al-2019_cvpr, liu-et-al-2021} suffer from artifacts in the same textureless regions as well, see Fig.~\ref{fig:real_overall} top.
Additionally, they generate ghosting artifacts for large motions, observable in Fig.~\ref{fig:eye}.
Overall, our method achieves a robust alignment and is able to construct thin structures like leaves, shown in Fig.~\ref{fig:real_overall} top and middle.

\subsection{Ablation Studies}
\label{sec:ablations}
To verify the effect of events in our pipeline, we evaluate \NAME~with and without event data input, which results in a pure image-based HDR method.
The image-based version of our architecture does not use the events in the feature alignment module and instead only uses the image features.
By including events, we see a performance boost of 5.4 dB, confirming the benefits of events.

This improvement can also be observed in the reconstructed HDR images, which contain more details than the image-only method, especially in objects, which are over-saturated in all the LDR brackets.
In Fig..~\ref{fig:wo_events}, the shape of the sun reflecting in water can only be inferred properly when modulation masks have access to events (via the proposed feature alignment module).
This shows that the high-dynamic-range of events is leveraged by the modulation masks of deformable convolutions to guide HDR generation in extreme conditions. 

To provide more insights on how events affect our method, we visualize the kernel offsets (DConv Offsets) (Fig.~\ref{fig:glass}) computed for the deformable convolutions.
By comparing the offsets between image and the combination of image and events, we see an enlarged receptive field visualized by the red circle on the glass falling down.
Thus, it can be concluded that the events improve the alignment by providing more accurate motion information.
Furthermore, we visualize the modulated mask for the deformable convolutions (DConv Mask). 
The mask predicted with events exhibits thin structure details, whereas the image-based uses a uniform weighting on the moving object.

Finally, we ablate the introduced network components by removing them from the architecture.
The results in Tab.~\ref{tab:hdm-hdr2014_ablations_short} show that each component improves the performance whereby the feature alignment has the largest impact.

\clearpage
\section{Conclusion}
\label{sec:conclusion}

We presented the first approach for multi-bracket HDR imaging with events. \NAME~ fuses motion information from images and events to enhance key parts of the HDR pipeline. 
As verified by our experiments, events significantly increase the performance on the real and synthetic data, confirming the robustness of our approach against misalignments.
Our approach also requires less f-stops to achieve the same performance as image-based alternatives. 
Finally, we recorded the first dataset that contains bracketed LDR images and synchronized events with HDR ground truth.

\section{Acknowledgement}
This work was supported by Huawei Zurich Research Center; NCCR Robotics, a National Centre of Competence in Research, funded by the Swiss National Science Foundation (grant number 51NF40\_185543); the European Research Council (ERC) under the European Union’s Horizon 2020 research and innovation programme (Grant agreement No. 864042).

\section*{\Large \bf Supplementary: Multi-Bracket High Dynamic Range Imaging with Event Cameras}
\section{f-stops Experiments on HDM-HDR-2014}

\newcommand\cellwidthhdmabl{1.7cm}

\begin{table*}
\caption{Quantitative comparison with state-of-the-art multi-bracket HDR approaches on the synthetic HDM-HDR-2014. The input LDR images are simulated with different f-stops to evaluate the performance with changing dynamic range in the input. Our methods outperforms all of the baselines, even showing a better performance compared to the baselines provided with more dynamic range information in the input. }
\vspace{-3mm}
\centering
\scalebox{0.9}{
\begin{tabular}{m{2.7cm}C{\cellwidthhdmabl}C{\cellwidthhdmabl}C{\cellwidthhdmabl}C{\cellwidthhdmabl}C{\cellwidthhdmabl}>{\centering\arraybackslash}m{\cellwidthhdmabl}}
  & \multicolumn{3}{c}{$\pm$2 f-stops} & \multicolumn{3}{c}{$\pm$4 f-stops} \\
 \cmidrule(lr){2-4} \cmidrule(lr){5-7} 
Method  & PSNR-$\mu$$\uparrow$ & SSIM-$\mu$$\uparrow$ & LPIPS$\downarrow$ & PSNR-$\mu$$\uparrow$ & SSIM-$\mu$$\uparrow$ & LPIPS$\downarrow$  \\
 \hline
Kalantari~\cite{kalantari-et-al-2017} & 31.34 & 97.41 & 0.056 & 40.16 & 98.23 & 0.029  \\
AHDR~\cite{yan-et-al-2019_cvpr}       & 39.03 & 98.76 & 0.029 & 39.95 & 98.68 & 0.021  \\
ADNet~\cite{liu-et-al-2021}           & 39.30 & 98.82 & 0.027 & 40.16 & 98.95 & 0.020  \\
 \hline
Ours                                  & \textbf{42.71} & \textbf{99.02} & \textbf{0.023} & \textbf{47.49} & \textbf{99.24} & \textbf{0.013} \\
\end{tabular}
}
\label{tab:exp_hdm-hdr_abl}
\end{table*}

Since we can synthetically create input LDR brackets with different exposure times on HDM-HDR-2014, we also evaluate the influence of different dynamic ranges in the input on the performance of all evaluated baselines.
Note that, we keep the same exposure value for the mid-exposed LDR image, and only modify the $\pm$ f-stops range for the short- and long-exposed LDR images.
Fig. 4 in the main manuscript gives a concise overview of the Table~\ref{tab:exp_hdm-hdr_abl}, which reports in addition to PSNR also the SSIM and LPIPS~\cite{Zhang18cvprLPIPS} metric.
It can be observed that our approach trained with only $\pm$2 f-stops range is outperforming the baselines trained with $\pm$4 f-stops range based on the SSIM and LPIPS metric as well.

\section{Limitations}
\begin{figure*}[h]
\centering
\includegraphics[width=0.97\textwidth]{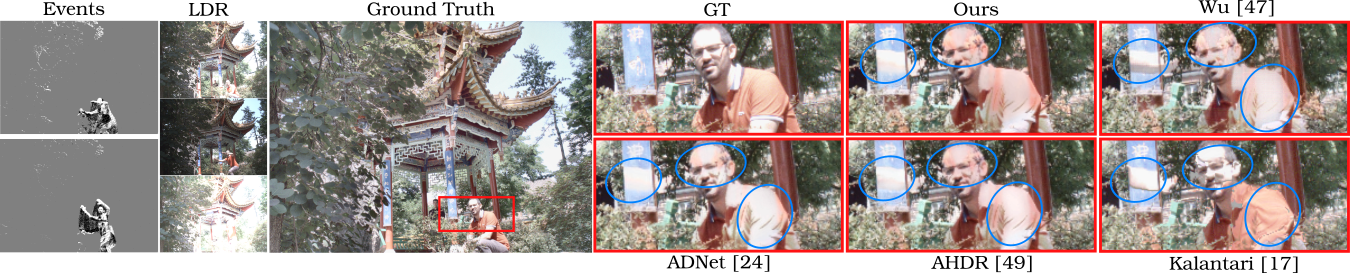}
\caption{Challenging example of our HDR-ERGB dataset. All of the tested methods struggle to correctly align the motion information from the LDR brackets to the reference frame. Nevertheless, our methods generate quantitatively the HDR image closest to the ground truth.}
\label{fig:standingup}
\end{figure*}

As can be seen in Fig.~\ref{fig:standingup}, our recorded HDR-ERGB dataset contains on purpose challenging samples with very large non-uniform motion.
In these samples, our proposed method still faces some challenges in aligning correctly the moving parts with the reference frame, partially due to saturation in the moving parts.
However, our methods still achieves the best results compared qualitatively to all of the tested state-of-the-art methods.

\section{Licenses for Code and Dataset}
As stated on the project page of HDM-HDR-2014 at \url{https://www.hdm-stuttgart.de/vmlab/hdm-hdr-2014/#FTPdownload}, the "Academic and educational use of the HdM-HDR-2014 data set is free".
In our code framework, we used several code packages which are either under MIT or BSD-2-Clause License. We refer to the submitted code for the corresponding source URL.

\section{HDR-ERGB Dataset}
\begin{figure*}[h]
\centering
\includegraphics[width=0.97\textwidth]{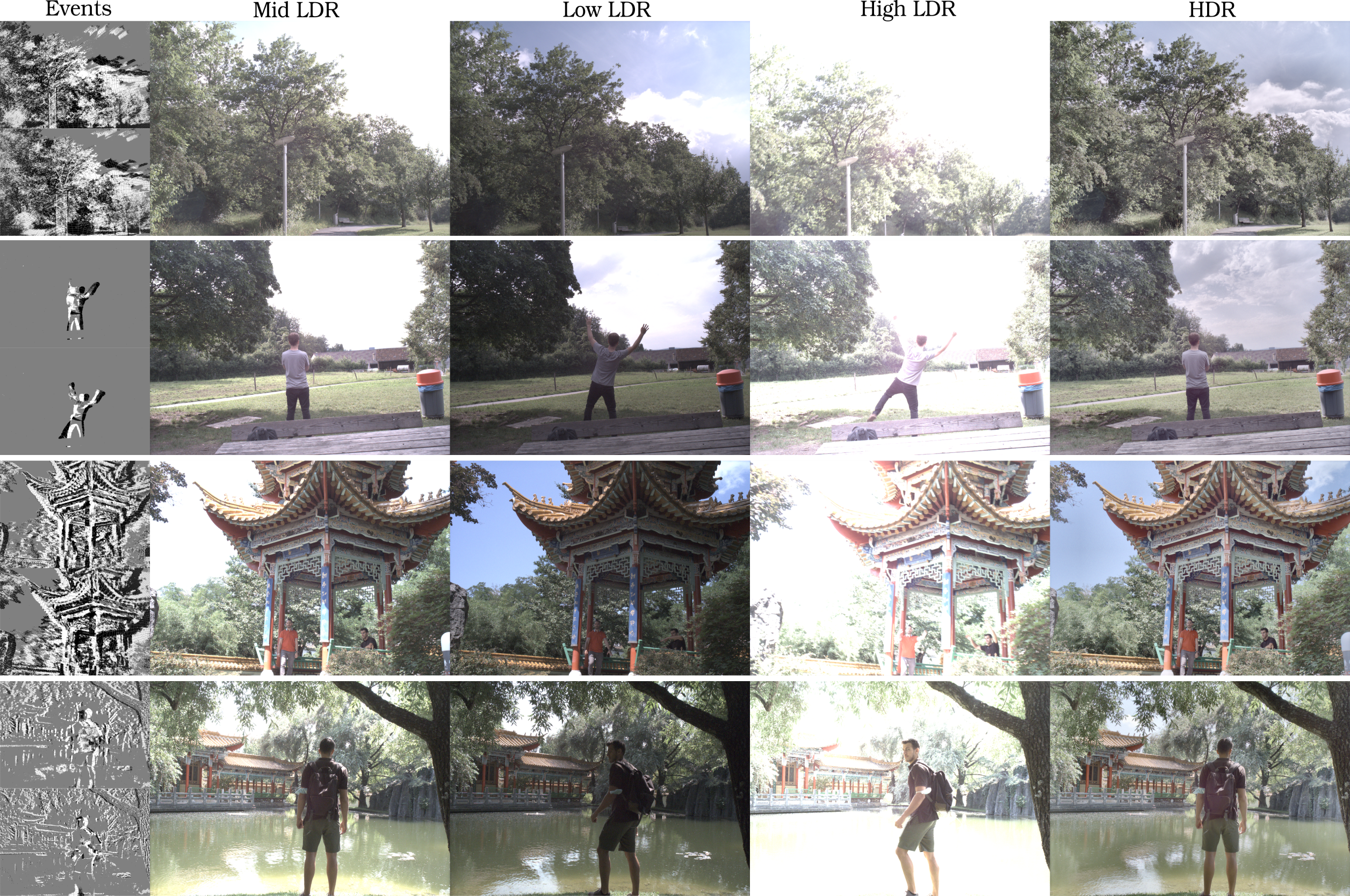}
\caption{Our recorded dataset includes samples containing only camera motion (top row), only scene motion (second top row) and combined camera and scene motion (third and last row).}
\label{fig:dataset_overall}
\end{figure*}

\begin{figure}[h]
\centering
\includegraphics[width=0.46\textwidth]{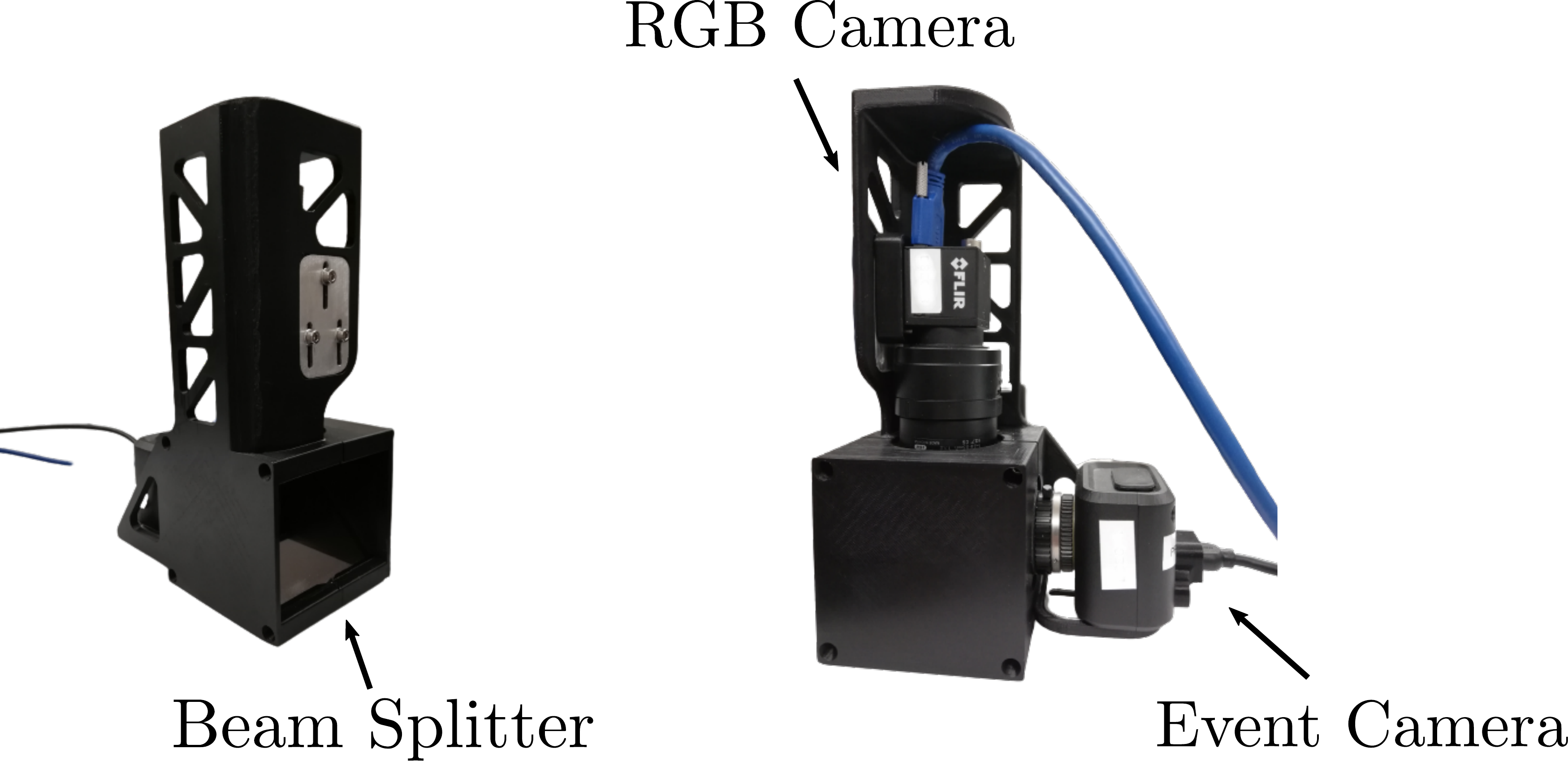}
\caption{The beamsplitter setup used to record our new HDR-ERGB dataset. It combines an event and RGB camera by projecting the scene via a beamplitter mirror to both cameras.}
\label{fig:beam_splitter}
\end{figure}

In this section, we describe in more detail how the sequences of our HDR-ERGB dataset were recorded using a beam splitter setup containing an event and RGB camera, see Fig~\ref{fig:beam_splitter}. We show several examples of our dataset in Fig.~\ref{fig:dataset_overall}. 

\subsection{Synchronization} 
In our beam splitter setup, the event and RGB cameras are hardware synchronized. Specifically, the RGB camera sends a trigger signal at the start and end of each exposure to our event camera, where those are recorded as external trigger events. These trigger events contain a precise timestamp for the clock time of the event camera, that allows to temporally synchronize the images with the event stream.

\subsection{Calibration} 
To calibrate the setup, we use the E2Calib toolbox~\cite{Muglikar2021CVPR}. We initially record a checkerboard pattern with both cameras. The toolbox then uses E2VID~\cite{Rebecq19pami, Rebecq19cvpr} to reconstruct intensity images from the asynchronous events, which are temporally synchronized with the standard frames. The calibration tool Kalibr~\cite{Oth_2013_CVPR, Maye-et-al-2013-Selfsupervised, Furgale-et-al-2013-kalibr, Furgale-et-al-2015-kalibr} is then used to obtain the intrinsics and extrinsics of both cameras. 
In a final step, the cameras are rectified using the camera parameters of the event camera.
\added{
In practice, the event and RGB camera still have a small parallax ($< \SI{1}{\milli\meter}$) in the z-direction of the corresponding camera coordinate system. 
However, since our HDR-ERGB dataset does not contain scenes recorded at a close distance to the camera, the stereo rectification provides pixel-accurate alignment. 
}

\subsection{Ground Truth HDR Acquisition}
\begin{figure}[h]
\centering
\includegraphics[width=0.47\textwidth]{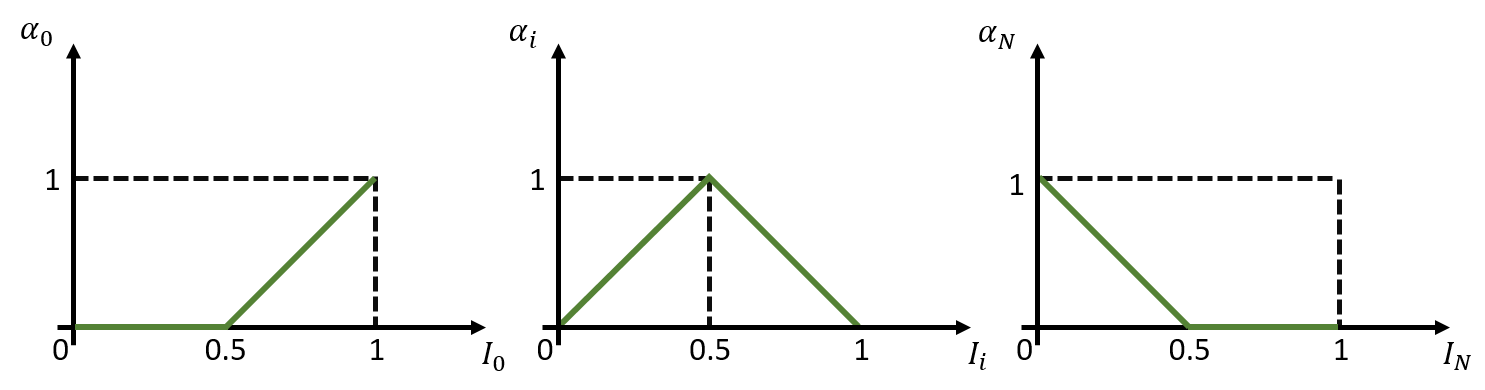}
\caption{The triangle functions used as weights $\alpha$ to generate the ground truth HDR image.}\label{fig:alpha_weights}
\end{figure}

As described in the paper, we record bracketed LDR images and events in two steps. In the fist step, we record a set of 9 bracketed LDR iamges on a steady tripod, which are then used to create the HDR ground truth. In step two, events and three LDR brackets are recorded containing camera and scene motion in form of moving persons. This way, the first LDR frame of the dynamic sequence is aligned with the ground truth HDR frame. In order to achieve this, a one second pause was introduced in between the first and second LDR image to allow the photographer and people in the shot to react.
To construct the HDR ground truth from the $N=9$ static LDR images, a simple triangle weighting scheme was adapted from~\cite{debevec-et-al-97} and~\cite{kalantari-et-al-2017} to merge the images. The LDR images were first linearized using the inverse camera response function and divided by the normalized exposure to map them into the HDR domain. $H_i = I_i^\gamma/t_i$ with $t_i$ being the exposure time of the image divided by the shortest exposure time in the sequence. In our notation the exposures are sorted from the shortest exposure time with index $i=0$ to the longest exposure time with index $i=N$. The images $H_i$ are then averaged with a weighted average scheme inspired by~\cite{debevec-et-al-97} and~\cite{kalantari-et-al-2017}. 
\begin{equation}
    H_{gt} = \frac{\sum_{i=0}^{N} \alpha_i(p) * H_i(p)}{\sum_{i=0}^{N} \alpha_i(p)}
\end{equation}
The weights $\alpha_i$ are obtained from the triangle functions depicted in Fig.\ref{fig:alpha_weights} evaluated on the $i$-indexed LDR image.


\clearpage

{\small
\bibliographystyle{ieee_fullname}
\bibliography{rpg,hdr}
}

\end{document}